\documentclass[twocolumn,showpacs,showkeywords,pre,amssymb]{revtex4}
\usepackage{amsmath}
\usepackage{bm}
\usepackage{graphicx}

\usepackage{setspace}
\begin{document}

\title{Nonlinear dynamics at the interface of two-layer stratified
flows over pronounced obstacles. }

\author{Cecilia Cabeza}
%\email{cecilia@fisica.edu.uy},
\author{Juan Varela}
%\email{jvarela@fisica.edu.uy},
\author{Italo Bove}
%\email{italo@fing.edu.uy},
\author{Daniel Freire}
%\email{dfreire@fisica.edu.uy},
\author{Arturo C. Mart\'{\i}}
%\email{marti@fisica.edu.uy}
\author{L. G. Saras\'ua}
%\email{sarasua@fisica.edu.uy}
\affiliation{Instituto de F\'{\i}sica,
  Universidad de la Rep\'ublica, Montevideo, Uruguay}

\author{Gabriel Usera}
%\email{gusera@fing.edu.uy},
\affiliation{Instituto de Mec\'anica de los Fluidos, Facultad de
   Ingenier\'{\i}a, Universidad de la Rep\'ublica, Montevideo, Uruguay}

\author{Raul Montagne\footnote{Present Address: Departamento de F{\'\i}sica
Universidade Federal Rural de Pernambuco
52171-900, Recife, PE, Brasil }
}

\author{Moacyr Araujo}
\affiliation{Laborat\'orio de Oceanografia F{\'\i}sica Estuarina e
Costeira, Departamento de Oceanografia, Universidade Federal de
Pernambuco, 50740-550, Recife, PE, Brazil}

\date{\today}

\begin{abstract}
The flow of a two--layer stratified fluid over an abrupt topographic
obstacle, simulating relevant situations in oceanographic problems, is
investigated numerically and experimentally in a simplified
two--dimensional situation. Experimental results and numerical
simulations are presented at low Froude numbers in a two-layer
stratified flow and for two abrupt obstacles, semi--cylindrical and
prismatic. We find four different regimes of the flow immediately past
the obstacles: sub-critical (I), internal hydraulic jump (II),
Kelvin-Helmholtz at the interface (III) and shedding of billows
(IV). The critical condition for delimiting the experiments is
obtained using the hydraulic theory. Moreover, the dependence of the
critical Froude number on the geometry of the obstacle are
investigated. The transition from regime III to regime IV is explained
with a theoretical stability analysis. The results from the stability
analysis are confirmed with the DPIV measurements.  In regime (IV),
when the velocity upstream is large enough, we find that
Kelvin-Helmhotz instability of the jet produces shedding of
billows. Important differences with flows like Von Karman's street are
explained. Remarkable agreement between the experimental results and
numerical simulations are obtained.
\end{abstract}


\maketitle

\section{Introduction}

The interaction between topography and flows with density interfaces is of interest both for fundamental reasons and for its relevance to
practical problems.  A most challenging problem is to describe and quantify the characteristic features occurring at the interface of a
stratified flow passing over an obstacle. The interaction between stable stratified flows and obstacles is a widespread phenomenon in
nature. In the atmosphere, for example, the flow around buildings or mountains is particularly important because such conditions are often
associated with high levels of atmospheric pollution due to low wind speeds and
suppressed vertical mixing \cite{baines,scinocca1989}. In physical oceanography,
the interaction of marine currents with topographic features, such as ocean
banks and coastlines, results in a complex system of circulation
\cite{scinocca1989,farmer99,Moum2000,Redondo2001,Torres2004,Klymak2004}. In this case, observational, analytical, numerical, and
previous laboratory studies
\cite{apsley95,travassos99,boyer00,taylor47,baines84,farmer86a,farmer86b, boyer87,Pawlak98,verron85,dewey05,Stastna2004,Stastna2005,Bonnier2000,bonneton00,baines2003,Jamali2005} suggest that the combination of streamlines splitting,
current intensification, and breaking of internal lee waves, play a significant
role as a mixing source in the ocean \cite{Ivey2008}.
Also, seamounts have a decisive impact on enhancing the biological productivity
and acting over the ecological
processes that determine the structure of local ocean life \cite{rogers94,genin04}. The rich diversity of geo-biophysical scenarios entirely justify the efforts for understanding the physics of the local instability at the density interface,  along with the conditions for which the generated lee waves break down.

Among the topographic effects on flows a particular problem that has received
considerable amount of attention is the oceanic observations in the Knight
Inlet, a fjord in central British Columbia, Canada.  The Knight Inlet
experiment provided field observations of tidal flow over a large sill
\cite{farmer99}. This work led to a series of papers reporting different issues
and possible explanation of them \cite{farmer99a,farmer2001,farmer2003,lamb04}.
This fjord has a strong tidal flow which generates internal
waves propagating along a pycnocline on both sides of
a sill there. Farmer and Armi \cite{farmer99} have observed that a very
large lee wave is formed behind the sill, while a train of
strong depression pulses are generated on the upstream side. 

Efforts has been dedicated to the upstream generation of solitary waves
\cite{farmer2001,Grue2005,Stastna2004,Stastna2005} and also on the 
trapped wedge of mixed fluid behind the sill. Upstream influence as a
consequence of variable forcing,  has been widely discussed in theoretical
analysis, laboratory experiments and numerical simulations
of stratified flow \cite{farmer02,farmer2003}. Flow separation by a topographic
obstacle  and the final stage of vortex shedding has also received attention
\cite{farmer99a,farmer2003,vosper99}.  Model simulations of the lee wave
formation by Lamb \cite{lamb04} exhibit a stable flow over the sill until the
lee wave overturns. All those work agree that the larger-scale response can be
sensitively dependent on small--scale instabilities. In the present work we focus on small--scale instabilities, addressing an analysis on the different regimes that appear downstream flow,  near the obstacle. The importance of the geometry of the  obstacle in the transition from subcritical to supercritical will be demonstrated. Finally, we prove the Kelvin--Helmholtz (KH) instability at the two layers interface, is the triggering  mechanism for the vortex shredding.

Under suitable conditions,  mainstream ceases to flow approximately parallel
to the obstacle beyond a certain point (as the flow in a divergent channel)
causing the phenomenon of separation \cite{Bat,lamb04}. In these cases, the flow
next to the wall usually reverses its direction in some regions.  Analytic
theories for lee waves and hydraulic jumps usually assume that flow separation
does not occur \cite{baines}. However, when the topography is abrupt, that is,
the vertical component of the velocity can not be neglected, flow separation is
expected. Thus, in the present paper we compare experimental and numerical
results with those obtained from hydraulic theories in the case of abrupt
obstacles.

The interaction between stratified flows and topography has been also investigated by means of laboratory experiments. In Ref.~\cite{bonneton00} the lee-wave breaking process which occurs at low Froude numbers has been studied in the case of a strong linear stratification and two-dimensional
smooth (Gaussian--shaped) obstacles. In a slightly different setup, the
structure of the far-wake vortices generated by a moving sphere
in a linearly stratified medium was investigated in Ref.~\cite{Bonnier2000}.  In this work the authors suggest that these vortices exhibit universal features similar to large-scale vortices found in the ocean.  The case in which both stratification and rotation are important has been study in Refs.~\cite{boyer87,boyer00}, where the role of laboratory experiments performed in close relation with numerical simulations is emphasized.

In our previous work \cite{varela07} we studied numerically and experimentally the different instabilities developed in a two--layer
stratified flow over a pronounced obstacle. The existence of a concentrated jet in the lee side of the obstacle was demonstrated and
we showed that the Kelvin-Helmholtz (KH) instability at the interface constitutes a more efficient source of mixing between the 
two--layers than the internal hydraulic jump. Here, we extend our work to two--layer flows over prismatic and semi--cylindrical abrupt 
obstacles focusing on the role of the geometry, the appearance of secondary instabilities after KH, and the shedding of billows.

The experiments were performed in a way that the flow could be assumed bi–dimensional.  We use two layers of different density, with constant
density inside each layer.  We look for different regimes of the flow past the obstacles. The experimental results, obtained via direct visualization and also with Digital Particle Image Velocimetry (DPIV), showed to be in qualitative good agreement with numerical simulations and theoretical results. Four different regimes were found, depending on the global Froude number $F_0$ and aspect ratio $H_m$.  A complete diagram in the parameter space was obtained. The transition from the regime III to IV was explained with a stability analysis of the jet past the obstacle. Those analytical results were validated by velocity measurements with DPIV technique.

This paper is organized as follows. In Section \ref{sec:modeldescrip} we briefly review the treatment of two--layer flows using hydraulic theory. 
In Section \ref{sec:experimentsetup} we present the experimental setup. The numerical simulations are given in Sec.~\ref{sec:numsim}. The results obtained from the experiments and simulations and their comparison are given in Sec.~\ref{sec:expresults}. In Section \ref{sec:staana} we present the
results of linear stability analysis, which are relevant for the flow past the obstacle. Finally, a summary and the conclusions are given in section \ref{sec:conclu}.

\section{Hydraulic theory of a  two--layer flow problem}
\label{sec:modeldescrip}

\begin{figure}
\begin{center}
\includegraphics[width=.6\columnwidth]{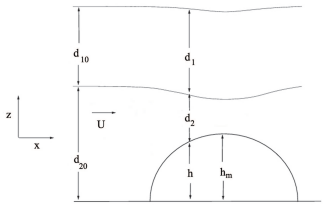}
 \end{center}
\caption{Sketch showing variables used in the description of the
two--layer flow. $\vec{U}$ is the upstream uniform fluid velocity,
$d_1$, $d_2$ are the depth of the layers and $h$ is the height of the
obstacle at a given location with $h_m$ the maximum height of it.}
\label{fig:sch}
\end{figure}

Let us consider a linear theory for a flow consisting of two layers of different
densities $\rho_1$ and $\rho_2$ ($\rho_2 >\rho_1$) over a fixed
obstacle.  Through this paper the subindexes $1$ and $2$ correspond respectively to the upper and lower layer.  We define a Cartesian
reference frame with coordinates $(x,y,z)$, where the flow is in the $x$ direction and $z$ is directed vertically upwards. As we mentioned in the introduction, we shall focus on the situations where variations in the $y$ direction can be neglected and the problem can be considered bi-dimensional. The depths of the layers are functions of $x$, as the thickness of them depends on the position where it is measured. Thus, the depth of the upper and lower layers are $d_1(x)$ and $d_2(x)$, respectively, and the height of the obstacle is $h(x)$ as sketched in figure \ref{fig:sch}.  Far upstream the depths are named $d_1(x=0) = d_{10}$ and $d_2(x=0)=d_{20}$.

The mean velocities of the fluid in each layer are $u_1$ and $u_2$. Let us assume that the fluid velocity is uniform far upstream,
with $u_{10}=u_{20}=U$.  In addition, we assume: a) the pressure is hydrostatic, b) the Boussinesq approximation, which implies that $\epsilon = (\rho_2 - \rho_1)/ \rho_2 \ll 1$, is valid, and c) the top boundary of the upper layer is a free surface at constant pressure $p_s$, taken to be $p_s=0$.  The flow over the topography is then characterized by the densities $\rho_1$, $\rho_2$, the depth of the
layers $d_1$, $d_2$, the mean velocity in each layer $u_1$, $u_2$ and the height of the obstacle $h$. With these assumptions, the Bernoulli
functions for each layer may be written as:
\begin{eqnarray}
E_1 = \rho_1 g (d_2+d_1+h) + \frac{1}{2} \rho_1 u_1^2  \nonumber  \\
E_2 = \rho_1 g d_1 +\rho_2 g (d_2 + h) + \frac{1}{2} \rho_2 u_2^2
\end{eqnarray}

Following Lawrence's model \cite{lawrence90,lawrence93}, from the conditions $dE_i/dx=0$ 
and imposing mass conservation in each layer the following relation is obtained
\begin{equation}
\frac{(1-F^2)}{\epsilon F_1^2 F_2^2} \frac{dD}{dx} = \frac{dh}{dx}
\label{G}
\end{equation}
where $F^2 = F_1^2 + F_2^2 - \epsilon F_1^2 F_2^2 $ is the composite
internal Froude number while $F_i^2=u_i^2 /(g' d_i)$ ($i=1,2$) are the
Froude numbers for each layer with $g'= (1 - \frac{\rho_1}{\rho_2})
g$, and $D = d_1 + d_2 + h$.  It has been shown \cite{lawrence90} that
$F$ is the adequate composite Froude number for characterizing a
two--layer flow; i.e. if $F > 1$ the flow is internally supercritical
(the internal small waves cannot propagate upstream against the
background flow), and if $F < 1$ the flow is subcritical (the
disturbances may propagate in both directions).  When $F=1$, the flow
is termed critical and this location is usually called a control point. The transition from subcritical to supercritical flow
is of special interest in our experiment. When the flow is supercritical, internal hydraulic jump may take place which is an
important source of turbulence and mixing. From Eq. (\ref{G}), it
follows that the critical condition $F = 1$, may occur if
$dh/dx=0$. For that value of $F$ the flow under goes a transition from
subcritical to supercritical when the surface has horizontal tangent,
i.e. at the crest of the obstacle. If, in addition, the condition
$\epsilon \ll 1$ is imposed (which is satisfied in our experiments),
the composite Froude number may be expressed as $F^2 = F_1^2+F_2^2$.
On the other hand, in the present study we consider flows where
$F_i^2 \lesssim 1$. Thus, from Eq.~(\ref{G}) and the
Boussinesq approximation it follows that
\begin{equation}
\frac{dD}{dx} =
 \frac{ \epsilon F_1^2 F_2^2}{ 1-F^2} \, \frac{dh} {dx} \approx 0 \, ,
\end{equation}
therefore we consider $D= d_{10}+ d_{20}$ a constant. Hence the free
surface will be taken as horizontal.

In order to obtain the critical values of the flow parameters for the
subcritical--supercritical transition, we impose that $E_2 - E_1$ is
constant and using the Boussinesq approximation, then
\begin{equation}
\frac{1}{2} r (1-r) F_0^2 \bigg( \frac{r_0^2}{r^2}-\frac{(1-r_0)^2}{(1-r_0
H - r)^2} \bigg) +r_0 (H-1)  +r  = 0
\label{Eq1}
\end{equation}
where $r = {d_{2}}/{D} $, $r_0 = {d_{20}}/{D} $, $H = {h}/{d_{20}} $ and
\begin{equation}
F_0 = \sqrt{\frac {U^2}{g'd_{10}}+\frac{U^2}{g'd_{20}}}
\label{eq:gfroude}
\end{equation}
is the Froude number $F$ calculated in the upstream flow far from the
obstacle. We shall call $F_0$ as the global Froude number.  On the
other hand, the critical condition $F^2 = 1$ may be expressed as
\begin{equation}
 r (1-r) F_0^2 \bigg( \frac{r_0^2}{r^3}+\frac{(1-r_0)^2}{(1-r_0 H -
 r)^3} \bigg) -1 = 0.
\label{Eq2}
\end{equation}

Thus, Eqs. (\ref{Eq1}) and (\ref{Eq2}), imposed at the crest of the
obstacle, may be used to determine, $F_{0c}$, the critical value of $F_0$ for the
occurrence of supercritical flow.
In our experiments  we fixed the aspect ratio $r_0 = 0.6$, 
and we considered different values of $H_m = h_m/d_{20} $,
$h_m$ being the height of the obstacle. 
The critical values,  $F_{0c}$, for the occurrence of critical flows as a function of $H_m$ are
obtained solving  equations (\ref{Eq1}) and (\ref{Eq2}).  The results are shown in
Sec.~\ref{sec:expresults}, Fig. \ref{fig:frdosbancos}, where are compared with the experimental results.

Theoretical solutions for the flow when $F_0$ is larger than $F_{0c}$, i.e. when
the flow is beyond the critical condition, have been obtained in Refs.\cite{baines,lawrence93}. 
These solutions predict that there is range of values of  $F_0$ 
for which there is a wave that moves backwards to the flow changing the conditions upstream. These approaches
allow the calculation of the velocity and amplitude of this wave.  This kind of wave has been also obtained  using the KdV equation \cite{Melville87}, and numerical simulations, with step like stratification \cite{Grue2005} and linearly varying stratification  \cite{Stastna2005}.
This effect may difficult the experiments inside the container in some cases. However, we
show that quantitative good results may be obtained within a broad range of parameter values.

\section{Experimental Setup}
\label{sec:experimentsetup}

\begin{figure}
\begin{center}
\includegraphics[width=.70\columnwidth]{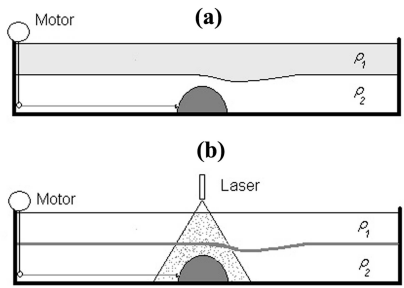}
\end{center}
\caption{Schematic diagram of the experimental configuration and the
two visualization techniques used. a) The upper layer is dyed with
$\mathrm{KMnO}_4$ to obtain good contrast for visualization when the
tank is lighted from behind. b) The DPIV is carried out lightening from
above with a LASER.  In this case, in order to distinguish the
interface, only a thin portion of the upper layer is dyed.}
\label{fig:setup}
\end{figure}

\begin{figure}
\begin{center}
\includegraphics[width=.49\columnwidth]{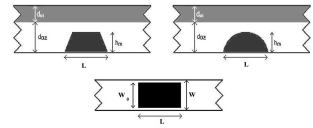}
 \end{center}
\caption{Side (top) and plan view (bottom) of the prismatic and
semi-cylindrical obstacles. Where $d_{10}$, $d_{20}$ are the depth of
the layers, $h_m$ the maximum height and $L$ the length of the
obstacle. $W$ is the width of the water tank and $W_0$ the width of
the obstacle.}
\label{fig:dimensiones}
\end{figure}

Our experiments were performed in a water tank of size equal to $2.0
\times 0.29 \times 0.137$ ${\mathrm m}^3$.  We used a closed channel
where we towed the obstacle at different velocities with a calibrated
motor (Fig.~\ref{fig:setup}).  The velocities of the fluid were
calculated in a reference frame fixed to the obstacle. In our previous
work \cite{varela07}, we showed that this configuration is equivalent
to the flow over a fixed obstacle. Moreover, we focused our
observations only on the central region of the tank where the
structures are persistent and the far boundaries effects can be
neglected.  In this work, we always show images and diagrams of
leftward moving obstacles.  We used two different obstacle shapes,
prismatic and semi--cylindrical, both of them having a height of
$h_m = 0.125$ m, width $W_0 = 0.13$ m and length of $L = 0.25$ m, see
Fig. \ref{fig:dimensiones}.  The obstacles were scaled in such a way
that the confinement aspect ratio verifies $W_0/W \lesssim 1$ ($W=
0.137$ m, width of the water tank) and the lateral flow around the
obstable can be neglected in order to reduce the problem to a
quasi-two dimensional situation.  As mentioned above, we want to model
a density profile with an abrupt gradient at the interface.  In order
to get this step-like stratification we first filled the tank with a
layer of density $\rho_2 = 1002$ ${\mathrm kg/m}^3$ using NaCl
solution.  To fill the upper layer, pure water with density $\rho_1 =
1000$ ${\mathrm kg/m}^3$ was carefully poured over a sponge floating
on the free surface.
The time scale of the molecular diffusion between the layers is much
longer than the typical experimental times.  Despite that, due to the
mixing produced by the moving obstacle, after a few measurements, the
tank had to be emptied and new fluid layers poured again.

The flow was examined via two standard techniques: dye visualization and
digital particle image velocimetry (DPIV).  In the first case, the upper layer
was dyed with a $\mathrm{KMnO}_4$ solution in order to obtain a good
visual contrast between both layers. A powerful source of fluorescent
light from behind was used to obtain a uniform illumination.  We
obtained global qualitative picture of the flow with this technique.
In the second case, DPIV allows us to obtain quantitative values of
velocity field based on the cross-correlation of two consecutive
images recorded by a digital camera. Neutrally buoyant polyamide
particles of $50\times10^{-6} \, m$ diameter were seeded in the bottom
layer. A green laser sheet of $100$ mW is used to illuminate a cross
section plane of the flow which is recorded by a digital camera
PIXELINK PL-A741.  In order to control the interface and avoid
attenuation of the LASER light as much as possible, when using this
technique, only a thin layer of water at the bottom of the upper layer
is dyed.  For both obstacle shapes, five different sets of heights
were chosen, always keeping constant the characteristic ratio $r_0 =
0.6$. The experiment was repeated for each set of heights with a wide
range of velocities analyzing the different behaviors, using the two
mentioned visualization techniques.

\section{Numerical Method}
\label{sec:numsim}

The numerical simulations considered here were obtained with the in-house flow
solver caffa3d.MB \cite{codeusera} developed jointly by Universitat Rovira i
Virgili (Tarragona, Spain) and Universidad de la Rep\'ublica
(Montevideo, Uruguay). It is an original Fortran95 implementation of a
fully implicit finite volume method for solving the 3D incompressible
Navier-Stokes equations in complex geometry using block structured
grids.  This three-dimensional solver, based on a previous two-dimensional
solver \cite{lilek97}, is described and validated in \cite{usera06,usera08}.

The unsteady incompressible Navier-Stokes equations with Boussinesq
approximation for buoyancy terms were considered. Since the Reynolds number was
below Re=$8.10^2$ for all cases, no turbulence model was required, thus
transient solutions were computed directly. The time step was set to $2.0 \times
10^{-2}$ s for all cases. This time scale is about $(h_m/U)/10^3$ for the
highest velocity case. Simulations were run starting from null velocity fields
through $10^4$ time steps, or about $200$~s of flow time.

In the simulations, the obstacle remains fixed against a steady two--layer
current of fluid. Thus a uniform velocity profile was specified at the upstream
boundary located at a distance of $8h_m$ upstream from the obstacle, and a null
gradient outlet was used at the downstream boundary, located $15h_m$ downstream.
As the top surface is not disturbed by the flow, it was modeled as an horizontal
slip boundary at fixed height. All other boundaries correspond to wall surfaces
and non-slip condition was directly applied to them, including the vertical
walls of the channel. Thus, the simulation accounts fully for three-dimensional
effects, although the flow reveals itself as essentially  two dimensional owing
to the geometry of the obstacles and the relatively low Reynolds numbers.

For both obstacles the grid was made up of three blocks, although the
topology was different in each case. For the prismatic obstacle three
straight blocks were assembled, two at each side of the obstacle and
the third extending on top of them along the domain.
On the other hand, for the cylindrical obstacle one C-grid block was used
around the obstacle together with two other straight blocks, upstream
and downstream of the obstacle.

Grid resolution was set essentially uniform through the domain at
$h_m/25$, being enough to resolve flow details at these rather
low Reynolds numbers. Due to the layout of the grid in the cylindrical
obstacle case the spatial resolution normal to the wall was slightly
higher near the obstacle, reaching about $h_m/35$.

\section{Results}
\label{sec:expresults}

\begin{figure}
\begin{center}
\includegraphics[height=11cm]{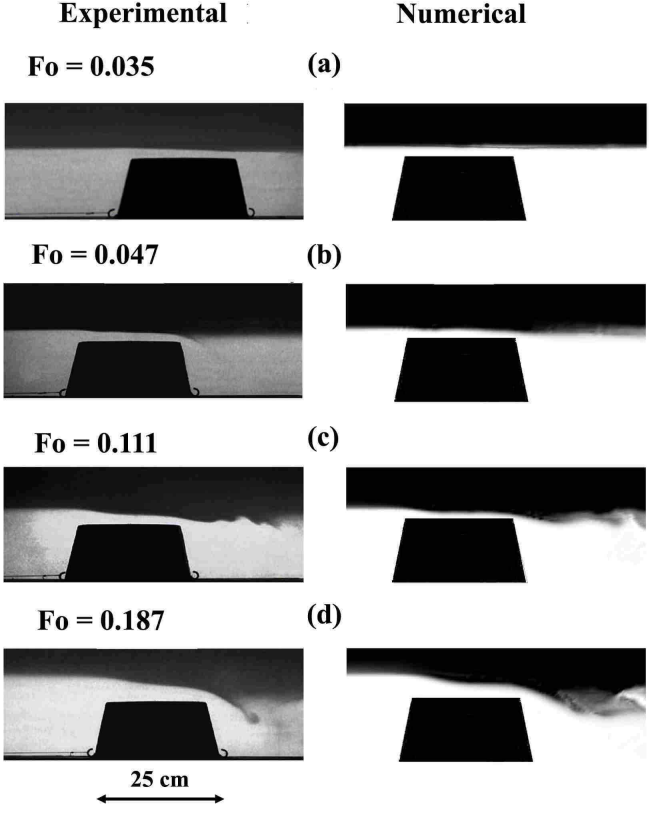}
 \end{center}
\caption{
Experimental snapshots (left) and numerical simulations of a flow (right) past an
prismatic obstacle corresponding to $d_{10} = 0.10$~m (upper layer), $d_{20} =
0.15$ m (bottom layer) corresponding to the different regimes. a) Subcritical regime (I). 
b) Internal hydraulic transition (II). c) KH instability at the interface (III). 
d) Billow formation (IV). 
In all the images the aspect ratio is $r_0 = 0.6$ and values of the the Froude numbers $F_0$ are indicated.}
\label{fig:figure4beta}
\end{figure}

\begin{figure}
\begin{center}
\includegraphics[height=14cm]{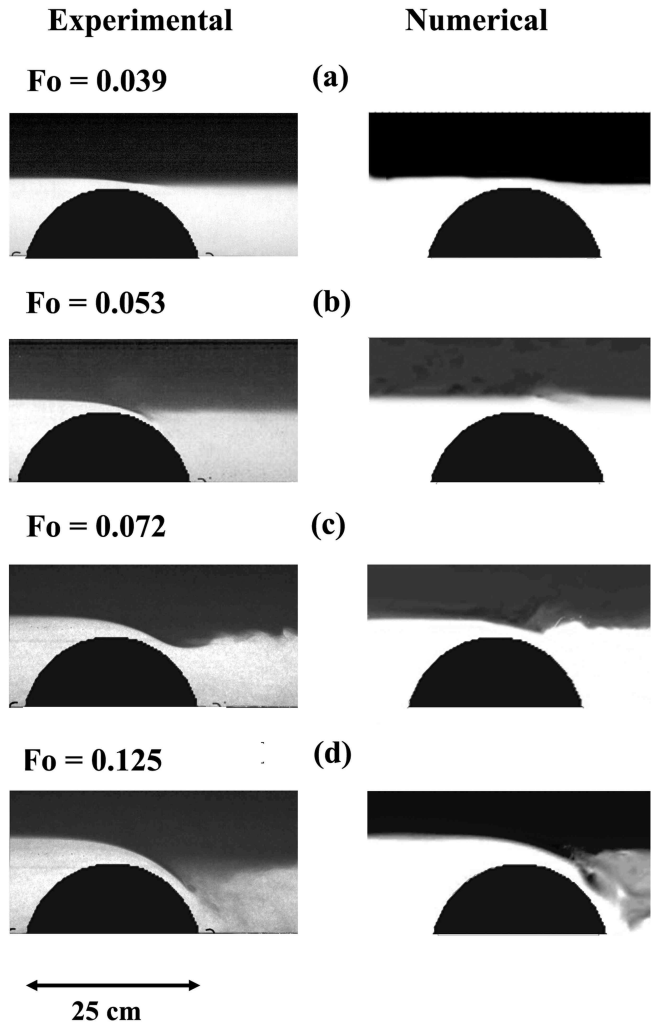}
 \end{center}
\caption{
 Experimental snapshots (left) and numerical simulations of a flow (right) past an semi-cylindrical obstacle 
corresponding to $d_{10} = 0.10$~m (upper layer), $d_{20} =0.15$ m (bottom layer) corresponding to the different regimes.
 a) Subcritical regime (I). b) Internal hydraulic transition (II). c) KH instability at the interface (III). d) Billow formation (IV).
Same parameters as in Fig.\ref{fig:figure4beta}.}
\label{fig:figure5}
\end{figure}

In this section we are going to discuss the experimental and numerical results,
which exhibit a great qualitative agreement  between them. It is also
worth noting that, in spite of quantitative differences, the results for both
topographic shapes show a qualitative similarity.
Let us start discussing the different regimes observed in the experiments. Using
the dye technique we visualized the different regimes as a function of the
obstacle velocity and the aspect ratio. In Fig.~\ref{fig:figure4beta}
(Fig.~\ref{fig:figure5}), we show experimental and numerical results for the
prismatic (semi-cylindrical) obstacle.  The aspect ratio and the height of the
bottom layer are the same for both obstacles, i.e. $r_0=0.6$ and $d_{20} = 15$ cm. 

In these figures we distinguish four regimes, all of them present in both
obstacles. However, both, experimental and numerical results reveals one very
important point: the regimes take place at different critical Froude number for
different geometries.  For the prismatic obstacle,  the velocities vary between
$0.12$ cm/s ($F_0 = 0.035$) and $0.64$ cm/s ($F_0 = 0.187$). While for the
semi-cylindrical obstacle, the velocities vary between $0.13$ cm/s ($F_0 =
0.039$) and $0.43$ cm/s ($F_0 = 0.125$).  The numerical simulation for both
obstacles were performed at the same global Froude numbers as the experimental
results. 

All the different behaviors of the downstream flow correspond to the case in which the flow upstream is subcritical, i.e.  $F_0 < 1$.  The thickness of the layers varies from one point to another immediately after the obstacle, as can be seen in Figs. \ref{fig:figure4beta} and \ref{fig:figure5}, and consequently the Froude number $F$ too.  We look for the first critical value of the local Froude number ($F=1$), over the obstacle in a place where $d_2$ is minimal and where the velocity is higher.

The regime (I) corresponds to the situations in which the flow is subcritical,
$F < 1$, along all the flow over the obstacle.  At low velocity, we observe
that a jet through the bottom layer is flowing near the obstacle. As the
velocity is increased, the jet begins to separate from the obstacle and rise
towards the horizontal. 

A transition from regime (I) to (II) occurs when Froude number reaches a control
point ($F=1$) somewhere over the obstacle. In regime (II) the flow is
subcritical upstream, and supercritical past the obstacle.  As a consequence, an
internal hydraulic transition is developed at the lee side.  The interface
between the layers is smoothly disturbed both in regime (I) and (II), however,
over the obstacle the interface is stable enough not to break but to induce a
jet in the lower layer.  When the velocity gradient between the jet
and the surrounding fluid is strong enough a Kelvin-Helmholtz instability
appears inside the lower layer at the lee side of the obstacle. The wavelength
of KH depends on two densities and the velocity profile (see Sec.~\ref{sec:staana}). The regimes (II) and on, are all  supercritical past the
obstacle,  $F >1$ beyond the critical point. 

Further increasing the Froude number $F_0$ (still below $1$), we reach regime
(III) where it is clearly visible a lee wave that perturbs the interface
separating the two layers, with a quasi-sinusoidal profile. Downstream, a
secondary instability develops and a kind of mixing is observed.  Finally, the
regime (IV) is characterized by the shedding of vortical portions of lighter
fluid separated from the upper layer. As a consequence, intense mixing between
the two layers takes place.  In this regime, the interface between both layers
is strongly disturbed. The jet drags fluid from the upper layer and billows
formation appear. It is worth mentioning that the frequency of the
shedding is almost constant with respect of variation in the velocity $U$. 

\begin{figure}
\begin{center}
\includegraphics[width=12.cm]{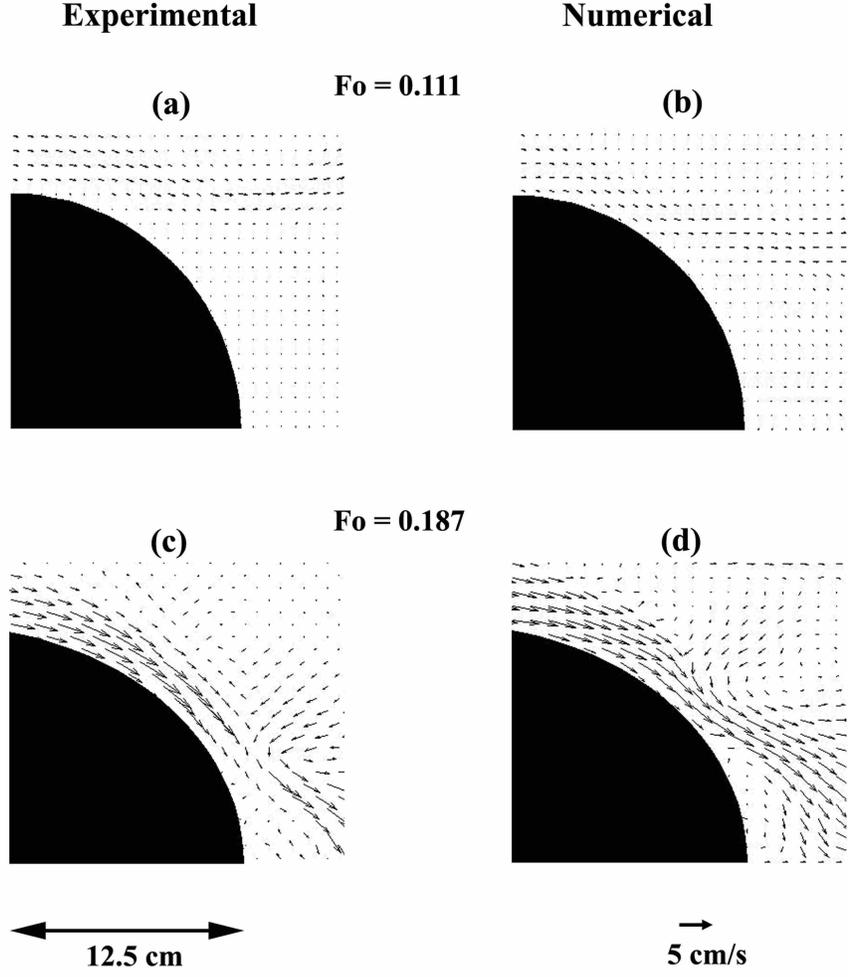}
\end{center}
\caption{Comparison between experimental (left) and numerical (right)
results for semi--cylindrical obstacle.  Top row $d_{20}=15$  cm
and $U=0.38$ cm/s and $F_0=0.111$. Bottom row $d_{20}=15$ cm and $U=0.64$ cm/s and $F_0=0.187$.  }
\label{fig:compara}
\end{figure}

The numerical results are in good agreement with the experimental results, as
can be visualized in Figs.~\ref{fig:figure4beta} and \ref{fig:figure5}.  The
regimes observed in the experiments were also obtained in the numerical
simulations for the two obstacles that we used.  As $F_0$ gets
larger the behavior of the numerical  and the experimental results differs.
For the case of the prismatic obstacle the vortex shedding starts  before
than in the simulations (Fig.~\ref{fig:figure4beta}d). Whereas in the cylinder
case the numerical experiments shows a vortex shedding before the experiments 
(Fig.~\ref{fig:figure5} d). These differences will be explained further on.
The overall agreement between numerical and experimental method is good as
can be seen in Fig. \ref{fig:compara} which compares numerical and experimental
results corresponding to regime III and IV. The
flow separation occurs before in the  experiments for low
$F_0$ (Fig.~\ref{fig:compara} a-b). However for greater $F_0$ the situation inverts
(Fig.~\ref{fig:compara} c-d). 

\begin{figure}
\begin{center}
\includegraphics[width=.60\columnwidth]{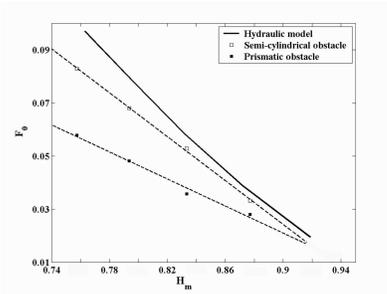}
 \end{center}
\caption{Critical Global Froude number as a function of the aspect
ratio $H_m = h_m/d_{20}$, for a fixed relation between the height of
the layers $r_0=0.6$.  The continuous line corresponds to the
hydraulic model and the symbols to the experimental results of the
two obstacles considered: semi-cylindrical (open squares) and prismatic
(full squares).}
\label{fig:frdosbancos}
\end{figure}

In Fig. (\ref{fig:frdosbancos}) we show the global Froude number $F_0$ as a function of the aspect ratio $H_m$ for the transition from
sub--critical to super--critical. The continuous line corresponds to the theoretical global Froude number for a smooth obstacle as it was
calculated solving equations (\ref{Eq1}) and (\ref{Eq2}). The experimental $F_0$ is represented with open squares for the
semi--cylindrical obstacle and filled squares for the prismatic one. The dashed lines are the linear fitting of the experimental
points. This graph clearly reflects the fact that the smoother the obstacle, the higher the velocity that will be needed to reach the
critical Froude number.  Then, the critical values for the semi--cylindrical obstacle are aligned between the values of the prismatic obstacle and the smoother theoretical prediction.  In this way, by means of the laboratory experiments we prove that he transition from one regime to another strongly depends on the geometry, the subcritical flow may correspond to a supercritical flow for another geometry.

\begin{figure}
\begin{center}
\includegraphics[width=.60\columnwidth]{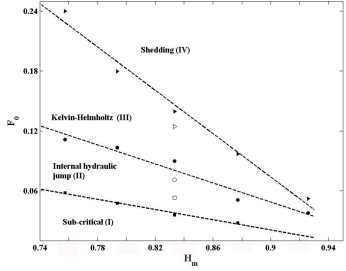}
\end{center}
\caption{
Stability diagram of the different regimes for the prismatic
and the semi-cylindrical obstacle as a function of the global Froude
number $F_0$ and the aspect ratio $H_m=h_m/d_{20}$. The relation between
the height of the two layers is fixed to $r_0=0.6$.  The full symbols
correspond to the experimentally obtained transition between different
regimes for the prismatic obstacle. Dotted lines are linear approximations of those experimental points. The open symbols
correspond to the transition points for the semi-cylindrical obstacle for
$H_m=0.83$. Squares (full and open), correspond to the transition between sub-critical to internal hydraulic jump regimen. Circles (full and open), correspond to the onset of Kelvin-Helmholtz regimen. Triangle symbol (full and open), correspond to the shedding of billows regimen.}
\label{fig:resumtinta}
\end{figure}

In order to obtain the transition between the different regimes, we use the DPIV technique. For the same parameters of the previous situation, $d_{20}= 15$ cm and prismatic obstacle, we observed that when the velocity is lower than $U = 0.11$ cm/s, we obtained a sub-critical flow. The transition between sub-critical to super-critical flow takes place at $U = 0.12$ cm/s ($F_0=0.035$).  For $
U = 0.38$ cm/s ($F_0=0.111$), KH instability at the interface
appears. Finally, for velocities higher than $U = 0.64$ cm/s ($F_0=0.19$) shedding of billows takes place.  Figure~\ref{fig:resumtinta} summarizes in a stability diagram the different regimes as a function of the global Froude number $F_0$ and
the aspect ratio $H_m$ for $r_0=0.6$. The diagram shows the transition
values between the regimes for the prismatic (full symbols) and semi-cylindrical (open symbols) obstacles.  The dotted lines are linear approximations. We show for $H_m = 0.83$, in open symbols the transitions between the different regimes for the cylindrical
obstacle.  It is interesting to note that, although the prismatic obstacle reaches the super critical regime at a lower velocity than
the semi-cylindrical obstacle, KH develops at higher velocities than in the case of the semi-cylindrical obstacle.

\begin{figure}
\begin{center}
\includegraphics[width=.70\columnwidth]{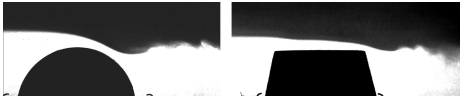}
\end{center}
\caption{Flow passing over the two obstacles of different
geometry. These snapshots correspond to the Kelvin-Helmholtz instability between both layers at $U=0.35$
cm/s for the prismatic obstacle and at $U=0.22$ cm/s for the
semi--cylindrical obstacle.}
\label{fig:compbancobeta}
\end{figure}

The slope of the jet after the obstacle plays an important role in the transition between the different regimes. Indeed, it was observed that transition depends not only on $F_0$, $r_0$ and $H_m$ but also in the shape of the obstacle, i.\ e.\ the slope of the jet after the obstacle. As a result, the lee waves past the prismatic and semi-cylindrical obstacles have similar characteristics at different Froude numbers. We exemplified this in figure \ref{fig:compbancobeta}. We show KH instability at the interface for the two obstacles. Despite the prismatic obstacle reaching the super critical regime at a lower velocity than the semi-cylindrical obstacle, the abrupt geometry of the bank makes difficult to rise up the jet and develop KH. Thus, for the prismatic obstacle, KH develops at higher velocities than in the case of the semi--cylindrical obstacle.

\begin{figure}
\begin{center}
\includegraphics[width=.98\columnwidth]{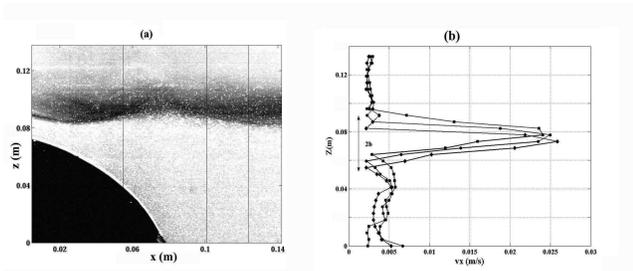}
\end{center}
\caption{ a)DPIV image for the semi-cylindrical obstacle for KH
instability at $F_0 = 0.122$. The vertical lines correspond to the places where
the velocity profiles were taken. b) Velocity profiles for the three
positions marked in the image displayed in  a)}
\label{fig:pivcirc}
\end{figure}

We characterized the velocity profile via DPIV analysis. Figure~\ref{fig:pivcirc} (b) shows the visualization of the
velocity profile over the semi--cylindrical obstacle obtained with DPIV
technique at the places indicated in Fig.~\ref{fig:pivcirc} (a).  We
observe that the shape of the jet is nearly triangular and quite constant at different distances from the obstacle. These facts will be
used to make an hydraulic stability analysis of the interface (see Sec.~\ref{sec:staana}).

In regime III as the flow velocity $U$ is increased, the inclination
of the jet, the amplitude of the interface wave and the velocity of
the jet increase. However, when the Froude number exceeds the critical
value, the wave at the interface is no longer observed and the flow
enters in regime IV. In this regime, there is shedding of vortical
portions of the more lighter fluid which submerge into the lower
layer.

\begin{figure}
\begin{center}
\includegraphics[width=.60\columnwidth]{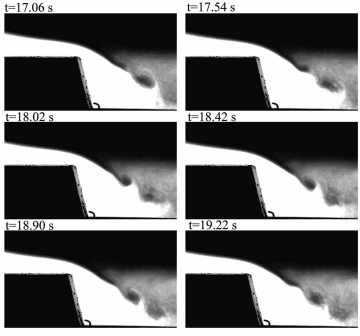}
\end{center}
\caption{Shedding of billows for $H_m=h/d_2=0.76$, $r_0=0.6$ and $F=0.26$
for the prismatic obstacle. The frequency of shedding is $0.5$ s$^{-1}$.}
\label{fig:iniciodespren}
\end{figure}

The shedding of billows in regimen (IV), as can be seen in figure \ref{fig:iniciodespren},
provide evidence that the instabilities which lead to this regimen could be caused by a KH instability.
In order to test this hypothesis, we correlate the thickness of the jet $a$ and the distance between the billows $x_b$. If the shedding was
caused by KH instability, the distance between the billows must be approximately the wavelength of the most unstable mode. Table I shows the values of $a$, $x_b$ and the ratio between them for the case $r_0 = 0.6$ and $H_m=0.83$, for the prismatic obstacle at different velocities.  These results reveals a clear proportionality between
$a$ and $x_b$, with a slope of $x_b/a = 2.7 $. Indeed, in the next section we consider a triangular jet, and we show that the
dimensionless wavenumber of the most unstable mode is given by $kb= 1.225$.  Then, the ratio between the corresponding wavelength and the
thickness of the jet $a$ ($a = 2b$) is given by $\lambda/a=2.57$, which is in very good agreement with the ratio $(2.6-2.7)$, obtained experimentally.

\begin{table}
\begin{center}
\begin{tabular}[t]{|c|c|c|c|}
\hline \ \ $U$ (cm/s) \ \ & \ \ $a$ (cm) \ \ & \ \ $x_b$ (cm) \ \ & \ \ $x_b/a$ \ \ \\ \hline 
0.45 & 2.1 & 5.4 & 2.6 \\ \hline 
0.55 & 2.3 & 6.2 & 2.7 \\ \hline 
0.65 & 2.8 & 7.4 & 2.6 \\ \hline
\end{tabular}
\caption{\label{Table} Thickness of the jet $a$ and distance between
the billows $x_b$ for $r_0=0.6$ and $H_m=0.83$.}
\end{center}
\end{table}

As commented in Sec.~\ref{sec:modeldescrip} theoretical solutions based on hydraulic theory predicts that for $F_0 > F_{0c}$
there is a wave that propagates in opposite sense to the flow, modifying the
upstream conditions. The role of this upsream wave has been profusely studied theoretical, observational and numerically
\cite{farmer99,farmer2001,farmer2003,lamb04}. We address here, the question of
how much this feature affects our results. 

\begin{figure}
\begin{center}
\includegraphics[width=.60\columnwidth]{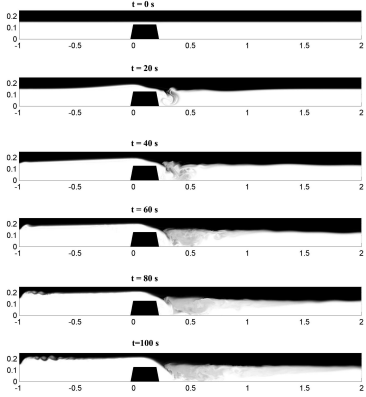}
\end{center}
\caption{Transient evolution of the interface showing the upstream influence of the wave corresponding to $F_0=0.35$.
The simulations start at $t=0$ with a null-velocity condition.}
\label{fig:upstream}
\end{figure}

In order to study the effect of this upstream wave, we performed
numerical simulations, in a broad range of values of the $F_0$,
focusing on the dynamics far upstream the obstacle and looking for the
transitory effects. The temporal evolution of the upstream wave, for
$F_0=0.35$, is shown in Fig.~\ref{fig:upstream}. We note that the
profile of the upstream wave is very smooth in comparison with the
downstream perturbation.  Accumulation of wave effects in the closed
container for Froude numbers clearly exceeding the critical condition
produces that the thickness of the layers are larger than those
obtained in an unlimited flow.  This effect is more important as
larger the $F_0$ is.  For instance, from the theory \cite{baines}
follows that for $F_0=0.181$ and $H_m= 0.833$, the increment of the
thickness is significant, $d_{10}'/d_{10} = 1.10$, where $d_{10}'$ is
the thickness of the interface taking into account the effects of the
upstream wave.  On the other hand, for $F_0=0.11$, $d_{10}'/d_{10} =
1.03$, this estimation shows that the effects of this wave are
negligible.  Furthermore we note that the results in
Fig.~\ref{fig:frdosbancos} and Fig.~\ref{fig:resumtinta} for
$F_0<0.11$ are not distorted in any way by the upstream effect.  In
the region $F_0>0.11$ altough the results could be slightly distorted,
they remains qualitatively correct as we can conclude from comparison
with the numerical simulations.

\section{Stability analysis of the jet past the obstacle}
\label{sec:staana}
\subsection{Kelvin-Helmholtz Instability}

As described in section \ref{sec:expresults}, in the lee side of the
obstacle the flow presents different regimes depending on the values
of $F_0$, $r_0$ and $H_m$.  However the type of flow downstream is not
only determined by the flow parameters but also by the slope of the
jet past the obstacle. We will analyze the stability of this flow.

In regime III, the interface between the two layers is perturbed by a
quasi--sinusoidal wave, which breaks down when its amplitude is
sufficiently large, indicating that a secondary instability takes
place. The DPIV measurements clearly show that the jet is located near
the interface, we conjecture that this wave results from a primary
Kelvin--Helmholtz instability. We test this hypothesis based in the
experimental measures shown in Fig. ~\ref{fig:pivcirc} (b) and we will
describe the flow near the interface as a triangular jet.  The flow is
given by
\begin{eqnarray}
V =\begin{cases}
             V_1 &  \text{for $z > b$ and $z < -b$}, \\
            V_2 + (V_1 - V_2) |z| /b  & \text{ for $ -b < z < b$} ,
         \end{cases}
\label{trijet}
\end{eqnarray}
where $2b$ is the thickness of the jet and the densities are $ \rho_2$
for $z<b$ and $\rho_1$ for $z>b$. We  perform  standard linear stability
analysis assuming a stream function of the perturbations
$\psi$  of the form \cite{drazin}

\begin{eqnarray}
\psi(x,z,t) = \phi(z) \exp(ik(x-ct)) \, ,
\end{eqnarray}
where $\exp(ik(x-ct))= \exp(ik(x-c_r t)) \exp{\sigma t} $, with
$\sigma = k c_i $, and $c_i, c_r$ are the imaginary and real parts of
the phase velocity $c$.  Thus, $\sigma$ represents the growing rate of
the perturbations. If $\sigma > 0$, then the flow is unstable to these
perturbations, and if $\sigma < 0$ the flow is stable.  Since the flow
under consideration has constant density in each layer, $\phi(z)$ is
determined by Rayleigh's equation:
\begin{eqnarray}
\phi''(z)-k^2 \phi(z)=0,
\end{eqnarray}
whose general solution is $\psi_i(z) = A_i \exp(k z) + B_i \exp(-k z)
$, where the index $i=1,2,3,4$ denotes the solution for the regions $z
> b$, $b > z > 0$, $0 > z > -b$ and $z < -b$ respectively. The perturbations
must be bounded at $z \rightarrow 
\infty$ and $z \rightarrow -\infty$, then the $\phi_i$'s become
\begin{eqnarray}
\phi_1(z) & = & B_1 \exp(-k z) \nonumber \\
\phi_2(z) & = & A_2 \exp(k z) + B_2 \exp(-k z) \nonumber \\
\phi_3(z) & = & A_3 \exp(k z) + B_3 \exp(-k z) \nonumber \\
\phi_4(z) & = &  A_4 \exp(k z).
\end{eqnarray}

We now impose the boundary conditions across the limits separating the
regions by requiring continuity of vertical velocity and
pressure. Since the velocity profile is continuous, the first of these
conditions is equivalent to require $\phi_i(z_j) = \phi_{i+1}(z_j)$
\cite{drazin}, where $z_j$ denotes the position of the boundary that
separates regions $i$ and $i+1$. The second of these conditions
requires
$$ (U_i(z_j) - c ) \phi'(z_j)_i - U_i'(z_j) \phi_i =
 (U_{i+1}(z_j) - c ) \phi'(z_j)_{i+1} - U_i'(z_j) \phi_{i+1}.$$

\begin{figure}
\begin{center}
\includegraphics[width=.49\columnwidth]{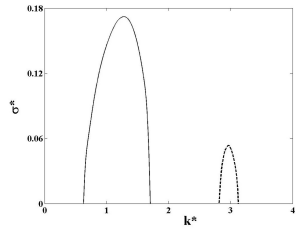}
\end{center}
\caption{Non-dimensional growing rate of the perturbations for the
unbounded stratified jet (continuous line) and the stratified jet near
a wall (dashed line), defined by Eqs.(\ref{trijet}) and
(\ref{trijet2}), where $k^* = kb$ and $\sigma^*= \sigma b /(V_2 -
V_1)$.}
\label{e-jet}
\end{figure}

The boundary condition yield a linear system of equation on $(B_1,
A_2, B_2, A_3, B_3, A_4)$.  The dispersion relation, i.\ e.\ the relationship
between $k$ and $c$, is obtained banishing the trivial solution. The
numerical results obtained by  solving the dispersion relation are shown in figure
(\ref{e-jet}). The normal mode stability analysis yields that the more unstable mode corresponds to the
wavenumber $k=1.225/b$.  For $b=1.1$ cm (see Fig. \ref{fig:pivcirc} b)
, the corresponding wavelength is $ \lambda =10.3$ cm which is in
good agreement with the experimental data of $\lambda=10.2$ cm. We
also performed the linear stability analysis of a jet that is located
near the wall of the obstacle, with field velocity given by
\begin{eqnarray}
V =\begin{cases}
             V_1 &  \text{for $z \ge b$}, \\
            V_2 + (V_1 - V_2) |z| /b  & \text{ for $-b < z < b$} ,
         \end{cases}
\label{trijet2}
\end{eqnarray}
where the rigid boundary is located at $z =-b$, and the densities are
$ \rho_2$ for $z<b$ and $\rho_1$ for $z>b$.  In this case, we have
three regions with corresponding stream functions $\phi_1(z),
\phi_2(z), \phi_3(z)$. The condition at the rigid boundary is given by
$\phi_3(-b)=0$ \cite{drazin}, and the other boundary conditions are
the same as the ones imposed for the unbounded jet. From this analysis we
obtained that the jet is partially stabilized in the proximity of the
wall of the obstacle, that is, the instabilities grow slowly compared
with the unbounded case (Fig. (\ref{e-jet})). This result is in
agreement with those obtained previously by Hazel \cite{hazel72}. This
stabilizing effect is caused by the combination of 
the density stratification and the presence of the wall.  Then the wall has
little effect on the stability  of the jet,  if the
density is constant.  This  explains why the instabilities appear
downstream, but not over the obstacle, where the jet is located on a
rigid boundary.

The stability analysis results and the measured wavelength of the lee
waves in the regimes III and IV, strengthen the assumption that the
instabilities observed past the obstacle in both cases are caused by
Kelvin--Helmholtz instability.  We can also explain the fact that the
frequency of the billows is almost independent of the velocity
$U$. From Table I we can see that the thickness of the jet is (for
fixed $r_0$ and $H_m$) practically proportional to $U$.  The frequency
of the billows can be expressed as $f = x_b / u_b$ , where $u_b$ is
the velocity of the billows. Since $x_b$ is nearly proportional to
$U$, if we assume that $u_b$ is proportional to $U$, then the
frequency is weakly dependent on the velocity $U$. This is in contrast
with vortices produced in flows like the Von Karman's street, where
the frequency, in a wide velocity range, is almost proportional to the
velocity.  The difference resides in the fact that in the Von Karman's
flow, the characteristic length is constant (the diameter of the
cylinder), while here the characteristic length (the thickness of the
jet) varies almost proportional with $U$, which causes that the
wavelength of the most dangerous mode increases with $U$. This effect
compensates the increasing of $u_b$ with $U$, leaving the frequency
almost constant.

\section{Summary and Conclusions}

\label{sec:conclu}

In this work, we studied two--layer stratified flow over abrupt obstacles, more specifically prismatic and semi-cylindrical.
It is remarkable that in all cases, the critical value of the global Froude number, $F_0$, is less
than that predicted by the hydraulic theory. In those cases, the vertical components of the velocity over the obstacles
cannot be neglected. As a consequence of its smoother geometry, the results for the semi-cylindrical obstacle are closer to the hydraulic
theory than those of the prismatic one. This result indicates that an abrupt obstacle reaches the control point for lower velocities
than for a smooth obstacle. In other words, for a given $F_0$,  a flow
subcritical for a geometry could be supercritical for another geometry.
  Despite the fact of being
approximate, the hydraulic theory allow us to estimate the thickness and
averaged
velocity of the strong jet that forms past the obstacle. However the
presence of flow separation and dissipation at the lee side of the
obstacle prevent the use of hydraulic theory for making further predictions.

For the downstream flow, four different regimes were identified and
represented in the parameter space $F_0 - H_m$. We showed typical
images of them in Fig.~\ref{fig:figure4beta}.  In all the regimes
observed the flow upstream remains subcritical. In regime I, the flow
is subcritical everywhere, i.e.\ upstream and downstream.  In regime
II and all the subsequent regimes the flow past the obstacle is
supercritical. The regime II is characterized by the transition from
subcritical to supercritical by the local Froude number experimentally
measured. Also in regime II, a shear instability develop mainly in the
lower layer. The interface that separates the layers is disturbed with
a quasi-sinusoidal wave in regime III. This disturbance is due to
Kelvin-Helmholtz instability. Furthermore, this quasi-sinusoidal wave
disturbance breaks down through a secondary instability for
sufficiently large amplitude and departure from two--dimensionality
takes place. Regime IV is characterized by the shedding of
billows. This process increases the rate of mixing between the two
layers.

A stability analysis on the jet immediately after the obstacle allowed
us to verify that the shedding was also produced by Kelvin-Helmholtz
instability. The characteristic length of the most dangerous mode of
the stratified jet is in agreement with the distance between the
billows.  We have also shown that the stabilizing effect resulting
from the combination of the proximity to an rigid boundary and
stratification turns out in not observing any instabilities over the
obstacle.

We studied the wave that propagates upstream when $F_0 > F_{0c}$. In the
simulations, we reproduced this wave with excellent agreement with the hydraulic
theory. The enhancement of the deeper layer thickness as well as the velocity of
the fluid inside it are in very good agreement with the theory \cite{baines}.
These results indicate that the effect of this wave  must be considered in flows
exceeding the critical conditions.  For $F_0$ of the order 0.1 or less, the
effects of the wave may be negligible. Furthermore, the results in
Fig.~\ref{fig:frdosbancos} and Fig.~\ref{fig:resumtinta} for $F_0<0.1$ are not
distorted in any way by the upstream effect. For this region of the space
diagram, and  high enough $H_m$, there are still the four different regimes, as
it was  described before and also in this  region vortex shedding takes place.
In the region $F_0>0.10$ the results remains qualitatively correct as we can
conclude from the numerical results comparison.

The experimental and simulation results showed that the inclination of
the jet plays a central role. Immediately after the separation, as velocity $U$
increases the inclination of the jet decreases.  However, when $U$ is
larger than a critical value, the inclination of the jet increases as
the velocity $U$ increases.  When the flow enters in regime IV, the
jet slope is pronounced. The different regimes reported in this work
appeared for both obstacles, but, since separation is a phenomenon
strongly dependent on the curvature of the surface,the critical values
for passing from one regime to another depend on the geometry of the
obstacle.

\section{Acknowledgments}
We acknowledge financial support from PEDECIBA (PNUD URU/06/004, Uruguay), Latin American Center for Physics (CLAF) and Grants PDT54/037. R.\ M.\ acknowledge financial support from the Brazilian agencies, CNPq, FAPERN, FACEPE.

%%%%%%%%%%%%%%%%%%%%%%%%%%%%%
%\bibliography{canal}

\begin{thebibliography}{45}
\expandafter\ifx\csname natexlab\endcsname\relax\def\natexlab#1{#1}\fi
\expandafter\ifx\csname bibnamefont\endcsname\relax
  \def\bibnamefont#1{#1}\fi
\expandafter\ifx\csname bibfnamefont\endcsname\relax
  \def\bibfnamefont#1{#1}\fi
\expandafter\ifx\csname citenamefont\endcsname\relax
  \def\citenamefont#1{#1}\fi
\expandafter\ifx\csname url\endcsname\relax
  \def\url#1{\texttt{#1}}\fi
\expandafter\ifx\csname urlprefix\endcsname\relax\def\urlprefix{URL }\fi
\providecommand{\bibinfo}[2]{#2}
\providecommand{\eprint}[2][]{\url{#2}}


\bibitem{baines}
P.~G. Baines,
  {\em Topographic Effects in Stratified Flows},
  Cambridge University Press, Cambridge, 1995.

\bibitem{scinocca1989}
J.F.~Scinocca and W.R.~Peltier,
 ``Pulsating downslope windstorm,''
J. Atmos. Scien. {\bf 46}, 2885 (1989).

\bibitem{Redondo2001}
J.A.~Carrillo, M.A.~Sanchez, A.~Platonov, and J.M.~Redondo,
``Coastal and interfacial mixing. Laboratory experiments and satellite observations,''
Phys. Chem. Earth B {\bf 26}, 305 (2001).

\bibitem{farmer99}
D.M.~Farmer and L.~Armi,
``Stratified flow over topography: the role of small-scale entrainment and mixing in flow establishment,''
Phil. Trans. R. Soc. Lond.  A {\bf 455}, 3221 (1999).

\bibitem{Klymak2004}
J.M.~Klymak and M.C.~Gregg,
  ``Tidally generated turbulence over the Knight inlet sill,''
  J. Phys. Ocean. {\bf 34}, 1135 (2004).

\bibitem{Moum2000}
J.N.~Moum and J.D.~Nash,
 ``Topographically induced drag and mixing at a small bank on the continental shelf,''
J. Phys. Ocean. {\bf 30}, 2049 (2000).

\bibitem{Torres2004}
C.R.~Torres, A.S.~{Mascarenhas Jr.}, and J.E.~Castillo,
  ``Three-dimensional stratified flow over Alarc\'on Seamount, Gulf of  California entrance,''
 Deep-Sea Research {\bf 11}, 647 (2004).

\bibitem{apsley95}
D.~Apsley and I.~Castro,
``Numerical computation of stratified flow over obstacles,''
in {\em Mixing in Geophysical Flows, International Center for Numerical Methods in Engineering},
edited by {J.\ M.\ }Redondo and O.~Metais, p.~87, Barcelona, Spain, 1995.

\bibitem{farmer86a}
L.~Armi and D.M.~Farmer,
  ``Maximal two-layer exchange through a contraction with a barotropic net flow,''
  J. Fluid Mech. {\bf 164}, 27 (1986).

\bibitem{baines84}
P.~Baines,
``A unified description of two--layer flow over topography,''
  J. Fluid Mech. {\bf 146}, 127 (1984).


\bibitem{baines2003}
P.G.~Baines and J.A.~Whitehead,
  ``On multiple states in single-layer flows,''
  Phys. Fluids {\bf 15}, 298 (2003).

\bibitem{Bonnier2000}
M.~Bonnier, O. Eiff, and P.~Bonneton,
``On the density structure of far-wake vortices in a stratified fluid,''
  Dyn. Atmos. Ocean {\bf 31}, 117 (2000).

\bibitem{boyer00}
D.~Boyer and P.~Davies,
``Laboratory studies of orographic effects in rotating and stratified flows,''
  Ann. Rev. Fluid Mech. {\bf 32}, 165 (2000).

\bibitem{boyer87}
D.~Boyer, P.~Davies, W.~Holland, F.~Biolley, and H.~Honji,
``Stratified rotating flow over and around isolated three-dimensional  topography,''
  Phil. Trans. R. Soc. Lond. A {\bf  322}, 213 (1987).

\bibitem{dewey05}
R.~Dewey, D.~Richmond, and C.~Garrett,
  ``Stratified tidal flow over a bump,''
  J. Phys. Ocen. {\bf 35}, 1911 (2005).

\bibitem{bonneton00}
O.~S. Eiff and P.~Bonneton,
``Lee--wave breaking over obstacles in stratified flow,''
Phys. Fluids {\bf 12}, 1073 (2000).

\bibitem{farmer86b}
D.M.~Farmer and L.~Armir,
``Maximal two-layer exchange over a sill and through the combination of a sill and contraction with a barotropic net flow,''
  J. Fluid Mech. {\bf 164}, 53 (1986).

\bibitem{Jamali2005}
M.~Jamali, B.~Seymour, and R.~Kasaiian,
``Numerical and experimental study of flow of a stratified fluid over a sill towards a sink,''
Phys. Fluids {\bf 17}, 057106 (2005).

\bibitem{Pawlak98}
G.~Pawlak and L.~Armi,
``Vortex dynamics in a spatially accelerating shear layer,''
J. Fluid Mech. {\bf 376}, 1 (1998).

\bibitem{Stastna2004}
M.~Stastna and W.R.~Peltier,
``Upstream-propagating solitary waves and forced internal-wave breaking in stratified flow over a sill,''
  Proc. R. Soc. Lond. A {\bf 460}, 3159 (2004).

\bibitem{Stastna2005}
M.~Stastna and W.~Peltier,
``On the resonant generation of large-amplitude internal solitary and solitary-like waves,
  J. Fluid Mech. {\bf 543}, 267 (2005).

\bibitem{taylor47}
G.~Taylor,
``Motion of solids in fluids when the flow is not irrotational,''
Phil. Trans. R. Soc. Lond. A {\bf 93}, 99 (1947).

\bibitem{travassos99}
P.~Travassos, F.~Hazin, J.~Zagaglia, R.~Adv{\'\i}ncula, and J.~Schober,
``Thermohaline structure around seamounts and islands off {N}orth--{E}astern {B}razil,''
  Arch. Fish. Mar. Res. {\bf 47}, 211 (1999).

\bibitem{verron85}
J.~Verron and C.~Le{P}rovost,
``A numerical study of quasigeostrophic flow over isolated topography,''
 J. Fluid Mech. {\bf 154}, 231 (1985).

\bibitem{Ivey2008}
G.N.~Ivey, K.B.~Winters, and J.R.~Koseff,
``Density stratification, turbulence, but how much mixing?''
Annu. Rev. Fluid Mech. {\bf 40}, 169 (2008).

\bibitem{rogers94}
{A.\ D.\ }Rogers,
``The biology of seamounts,''
Adv. Mar. Biol. {\bf 30}, 305 (1994).

\bibitem{genin04} A.~Genin,
``Bio-physical coupling in the formation of zooplancton and fish aggregations over abrupt topographies,''
  J. Mar. Sys. {\bf 50}, 3 (2004).

\bibitem{farmer2003}
P.~F. Cummins, S.~Vagle, L.~Armi, and D.M.~Farmer,
``Stratified flow over topography: upstream influence and generation of nonlinear internal waves,''
   Phil. Trans. R. Soc. Lond. A {\bf  459}, 1467 (2003).

\bibitem{farmer99a}
D.M.~Farmer and L.~Armi,
``The generation and trapping of solitary waves over topography,''
Science {\bf 283}, 188 (1999).

\bibitem{farmer2001}
D.M.~Farmer and L.~Armi,
``Stratified flow over topography: models versus observation,''
Proc. R. Soc. Lond. A {\bf 457}, 2827 (2001).

\bibitem{lamb04} K.~G. Lamb,
``On boundary-layer separation and internal wave generation at the {K}night {I}nlet sill.''
Phil. Trans. R. Soc. Lond. A {\bf 460}, 2305 (2004).

\bibitem{Grue2005}
J.~Grue,
``Generation, propagation, and breaking of internal solitary waves,''
  Chaos {\bf 15}, 037110 (2005).

\bibitem{farmer02}
L. Armi and D.M.~Farmer
``Stratified flow over topography: bifurcation fronts and transition to uncontrolled state,''
Phil. Trans. R. Soc. Lond. A {\bf  458}, 513 (2002).

\bibitem{vosper99}
S.~Vosper, I.~Castro, W.~Snyder, and S.~Mobbs,
``Experimental studies of strongly stratified flow past three--dimensional orography,''
  J. Fluid Mech. {\bf 390}, 223 (1999).

\bibitem{Bat}
G.~Batchelor,
  {\em An introduction to fluid dynamics},
  Cambridge University, New York, eighth edition, 2007.

\bibitem{varela07}
J.~Varela, M.~Aaujo, I.~Bove, C.~Cabeza, G.~Usera, A.~C. Mart{\'\i}, R.~Montagne, and L.~Saras{\'u}a,
``Instabilities developed in stratified flows over pronounced obstacles,''
Physica A {\bf 386}, 681 (2007).

\bibitem{lawrence90}
G.~A. Lawrence,
``On the hydraulics of boussinesq and non-boussinesq two-layer flows,''
  J. Fluid Mech. {\bf 215}, 457 (1990).

\bibitem{lawrence93}
G.~A. Lawrence,
``The hydraulics of steady two-layer flow over a fixed obstacle,''
J. Fluid Mech. {\bf 254}, 605 (1993).

\bibitem{Melville87}
W.K.~Melville and K.R.~Helfrich,
``Transcritical two-layer flow over topography,''
J. Fluid Mech. {\bf 178}, 31 (1987).

\bibitem{codeusera}
G.~Usera, the code is freely available for academic use through the web site
  (www.fing.edu.uy/imfia/caffa3d.MB).

\bibitem{lilek97}
Z.~Lilek, , S.~Muzaferija, M.~Peric, and V.~Seidl,
``Computation of unsteady flows using non-matching blocks of structured grid,''
  Numerical Heat Transfer. Part B, Fundamentals {\bf 32}, 403 (1997).

\bibitem{usera06}
G.~Usera, A.~Vernet, and J.~Ferr{\'e},
``Use of time resolved {PIV} for validating {LES/DNS} of the turbulent
 flow within a {PCB} enclosure model,''
 Flow Turbulence Combust. {\bf 77}, 77 (2006).

\bibitem{usera08}
G.~Usera, A.~Vernet, and J.~Ferr{\'e},
``A parallel block-structured finite volume method for flows in complex geometry with sliding interfaces,''
 Flow Turbulence Combust. (in press), (2008).

\bibitem{drazin}
P.~Drazin and {W. H.}Reid,
 {\em Hydrodynamic Stability},
 Cambridge University Press, Cambridge, second edition, 2004.

\bibitem{hazel72}
P.~Hazel,
``Numerical studies of the stability of inviscid stratified flows,''
 J. Fluid Mech. {\bf 51}, 39 (1972).

\end{thebibliography}

%%%%%%%%%%%%%%%%%%%%%%%%%%%%%

\newpage 

\end{document}